\documentclass[sigchi]{acmart}

\usepackage{booktabs} 

\usepackage{balance}       
\usepackage{graphics}      
\usepackage{url}      

\usepackage{csquotes}
\usepackage{enumitem}

\usepackage{url}      
\usepackage{color}
\usepackage{siunitx}
\newcolumntype{L}[1]{>{\raggedright\let\newline\\\arraybackslash\hspace{0pt}}m{#1}}
\newcolumntype{C}[1]{>{\centering\let\newline\\\arraybackslash\hspace{0pt}}m{#1}}
\newcolumntype{R}[1]{>{\raggedleft\let\newline\\\arraybackslash\hspace{0pt}}m{#1}}

\newenvironment{myquote}%
  {\list{}{\leftmargin=0.2in\rightmargin=0.2in}\item[]}%
  {\endlist}

\copyrightyear{2019}
\acmYear{2019}
\setcopyright{acmlicensed}
\acmConference[CHI 2019]{CHI Conference on Human Factors in Computing Systems Proceedings}{May 4--9, 2019}{Glasgow, Scotland Uk}
\acmBooktitle{CHI Conference on Human Factors in Computing Systems Proceedings (CHI 2019), May 4--9, 2019, Glasgow, Scotland Uk}
\acmPrice{15.00}
\acmDOI{10.1145/3290605.3300468}
\acmISBN{978-1-4503-5970-2/19/05}

\begin{document}
\title[Unremarkable AI]{Unremarkable AI: Fitting Intelligent Decision Support \\into Critical, Clinical Decision-Making Processes}

\author{Qian Yang}
\affiliation{%
  \institution{HCI Institute}
  \institution{Carnegie Mellon University}
}
\email{yangqian@cmu.edu}

\author{Aaron Steinfeld}
\affiliation{%
  \institution{Robotics Institute}
  \institution{Carnegie Mellon University}
}
\email{steinfeld@cmu.edu}

\author{John Zimmerman}
\affiliation{%
  \institution{HCI Institute}
  \institution{Carnegie Mellon University}
}
\email{johnz@cs.cmu.edu}

\renewcommand{\shortauthors}{Q. Yang et al.}

\begin{abstract}
Clinical decision support tools (DST) promise improved healthcare outcomes by offering data-driven insights. While effective in lab settings, almost all DSTs have failed in practice. Empirical research diagnosed poor contextual fit as the cause. This paper describes the design and field evaluation of a radically new form of DST. It automatically generates slides for clinicians' decision meetings with subtly embedded machine prognostics. This design took inspiration from the notion of \textit{Unremarkable Computing}, that by augmenting the users' routines technology/AI can have significant importance for the users yet remain unobtrusive. Our field evaluation suggests clinicians are more likely to encounter and embrace such a DST.
Drawing on their responses, we discuss the importance and intricacies of finding the right level of unremarkableness in DST design, and share lessons learned in prototyping critical AI systems as a situated experience.
\end{abstract}
%
%



\maketitle

\section{Introduction}

The idea of leveraging machine intelligence in healthcare in the form of decision support tools (DSTs) has fascinated healthcare and AI researchers for decades. These tools often promise insights on patient diagnosis, treatment options, and likely prognosis. With the adoption of electronic medical records and the explosive technical advances in machine learning (ML) in recent years, now seems a perfect time for DSTs to impact healthcare practice.

Interestingly, almost all these tools have failed when migrating from research labs to clinical practice in the past 30 years \cite{elwyn2013many,jaspers2011effects,kawamoto2005improving}. In a review of deployed DSTs, healthcare researchers ranked the lack of HCI considerations as the most likely reason for failure \cite{Musen2014clinical,wears2005computer}. This includes a lack of consideration for clinicians' workflow and the collaborative nature of clinical work. The interaction design of most clinical decision support tools instead assumes that individual clinicians will recognize when they need help, walk up and use a system that is separate from the electronic health record, and that they want and will trust the system's output.

We are collaborating with biomedical researchers on the design of a DST supporting the decision to implant an artificial heart. The artificial heart, VAD (ventricular assist device), is an implantable electro-mechanical device used to partially replace heart function. For many end-stage heart failure patients who are not eligible for or able to receive a heart transplant, VADs offer the only chance to extend their lives. Unfortunately, many patients who received VADs die shortly after the implant~\cite{benza2012evaluation}. In this light, a DST that can predict the likely trajectory a patient will take post-implant, should help identify the patients who are mostly likely to benefit from the therapy. 

We draw insight from a field study investigating the VAD decision processes, searching for opportunities where ML might help \cite{yangVAD2016}. The findings revealed that clinicians are unlikely to encounter or to actively engage with a DST for help at the time and place of decision making. For most cases, they did not find the implant decision challenging; thus, they had no desire for computational support. In addition, the extremely hierarchical healthcare culture stratified senior physicians who make implant decisions and the mid-level clinicians who use computers. Almost no VAD decision-making took place in front of a computer.

Embracing the rich context of the implant decision, we designed a radically new DST that automatically generates slides for the required decision meeting. The design embeds prognostic decision supports into the corner of decision meeting slides. We wanted decision makers to encounter the computational advice at a relevant time and place across the decision process, and we wanted this support to only slow them down for the few cases where the DST adds value to the decision. This design draws inspiration from Tolmie et al.'s notion of Unremarkable Computing, that technology needs to have the right level of remarkableness to valuably situate itself in people emerging routines and becoming the glue of their everyday lives \cite{unremarkable-computing}.

This paper presents this DST's interaction design as well as a field evaluation at three VAD implant hospitals. We also spoke with physicians working on clinical decisions outside of VAD implant, probing whether this design might generalize to other critical, clinical decisions. Our findings suggest that clinicians are more likely to encounter and embrace a DST that binds ``unremarkable" decision supports with their current work routine. Drawing on clinicians' responses, we discuss the importance and intricacies of finding the right level of unremarkableness in a DST design. We discuss lessons learned and unexpected challenges in evaluating critical AI systems as a situated experience. 

This paper makes two contributions. First, we offer one concrete solution to the long-standing challenge of effectively situating DSTs in clinical practice. Second, we offer a rare description of clinicians' responses to a DST situated in their workflow. This surfaced intriguing insights valuable for future investigations of critical AI systems.

\section{Related Work}

\subsection{Clinical Decision Support Tools in Practice}

Clinical decision support tools (DSTs) are computational systems that support one of three tasks: diagnosing patients, selecting treatments, or making prognostic predictions of the likely course of a disease or outcome of a treatment \cite{yang2015review}. 

This project focuses on clinician-facing, prognostic DSTs. A significant strand of recent HCI work focused on critical issues in this area, including AI interpretability and fairness, data visualization, accuracy of risk communication, and more \cite{tait2010effect,timmermans2004different,Clinical-viz}. The significance of this body of work has led some to describe it as ``the rise of design science in clinical DST research" \cite{rise-of-design-science}. These studies typically investigated DST in lab settings, using prototypes that are dedicated to a single clinical decision. Clinicians came out of their day-to-day workflow, used these systems for a pre-identified task, then provided feedback on the system design. 

Despite success in labs, the vast majority of clinician-facing DSTs failed when moving to clinical practice. Clinicians rarely use them \cite{devaraj2014barriers,elwyn2013many,wyatt1995commentary}, and therefore they have shown no significant improvement on healthcare outcomes. Healthcare researchers identified a lack of HCI consideration, rather than poor technical performance, as the main cause of these failures \cite{sittig2008grand,taneva2014meaning}. These HCI considerations include workflow integration, integration with social context, and clinicians' lack of motivation to use a DST. 

Relatively few research projects have focused on how to integrate DST output into clinical contexts. Within HCI there are initiatives to engage with real clinical users and contexts, yet the lack of meaningful access remains a major barrier. Researchers have shared that they were not allowed to evaluate incomplete designs, that evaluations took months or even years, that iterative design or repeated evaluation was not possible, and finally, that design evaluations had to be conducted by healthcare professionals rather than by HCI professionals~\cite{design-in-med,chi-healthcare-workshop,difficult-clinical-design-eval}.

\subsection{VAD Decision-Making and Its Context}
Due to many of the aforementioned barriers, investigations of the VAD implant decision making and field assessment of DST designs are rare.
An exception is a field study conducted at three VAD programs \cite{yangVAD2016}. Researchers made a number of observations that informed this work:

First, clinicians perceived no need for computational support; They considered most patient cases as textbook cases that follow a standard, systematic process of therapy escalation and a staged unfolding of decision considerations. 

Second, clinicians made implant decisions during daily rounding of patient wards, during hallway conversations, and in multidisciplinary VAD decision meetings. Decisions were rarely discussed or made in front of a computer.

Finally, the clinical workplace culture was strongly hierarchical yet highly collaborative. Cardiologists and surgeons, who function at the top of the hierarchy, decided who gets classified as a difficult case and who gets discussed during the required multidisciplinary meeting which the whole VAD team attends. This cultural context poses a two-fold challenge for DST use. First, decision makers (physicians) and computer users (the midlevel clinicians, including nurse practitioners, social workers and VAD coordinators) rarely overlap at any point of the decision-making process. Second, physicians have great trust in their colleagues' suggestions, much more so than in computational support.

\section{Design Process and Rationale}

We set out to design a new form of DST for VAD patient selection to explore how to overcome its real-world adoption barriers that many prognostic DSTs face. Drawing upon prior work, we had two design goals:

1 - \textit{Embedding DST in current workflow}: Clinicians, especially cardiologists and surgeons, need to naturally encounter the DST within their current decision-making workflow, because they are unlikely to recognize when they might need help and then walk up to a computer for help;

2 - \textit{Slowing down decision-making only when necessary}: The DST outputs need to be easily ignored in most patient cases that are textbook. However, it should also be present enough to slow the decision-making down when there is a meaningful disagreement between the clinicians' view and the DSTs view of the situation;

These orientations are very different from the convention of DST design in which decision supports are always available, waiting for clinicians to walk up and use at any point across the decision-making process. Instead, we wanted to tailor the DST for particular moments in the process, such that clinicians do not have to take pause and invent sequences of action anew. We wanted the DST to naturally augment the actions of decision making, rather than pulling the user away from doing their routine work.

\begin{figure*}
    \centering
    \includegraphics[width=0.72\textwidth]{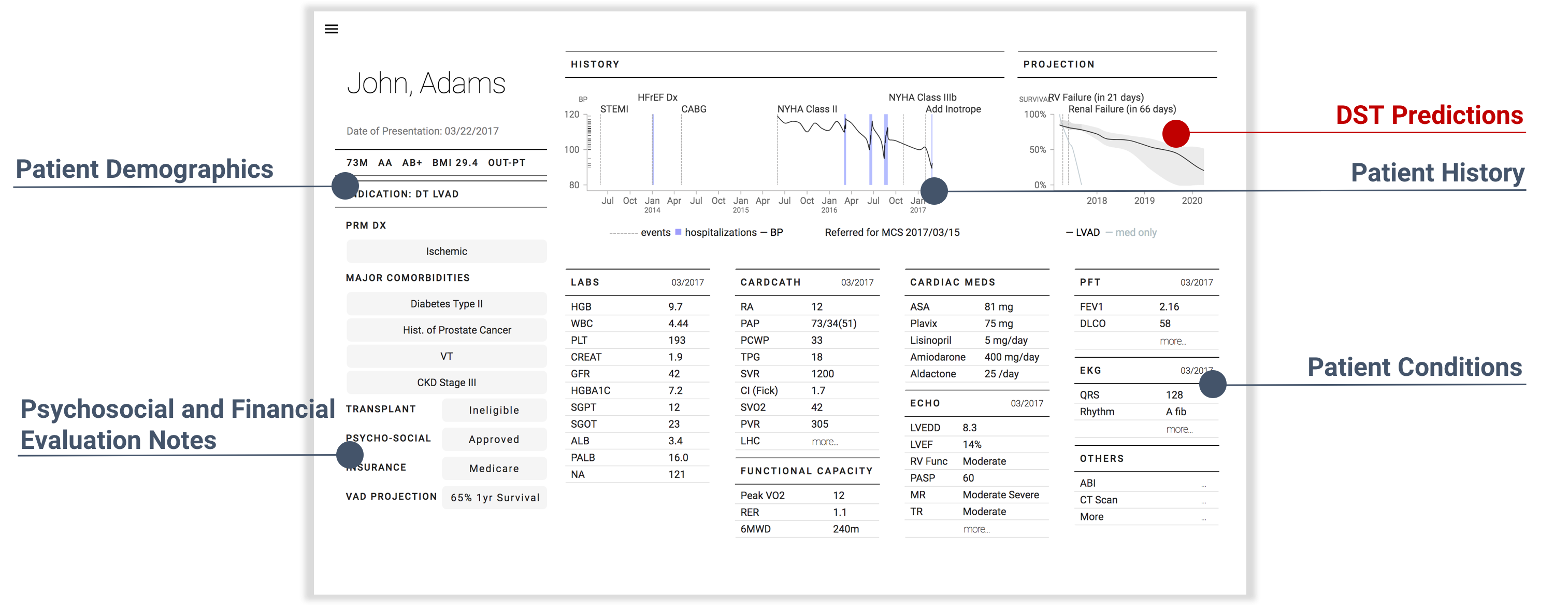}
    \caption{The decision meeting slide design. We designed a DST that automatically generates decision-meeting slides for clinicians with subtly embedded machine prognostics at the top right corner.}
    \vspace{0.05cm}
    \label{fig:screenshot}
\end{figure*}

\subsection{Making Clinical DST Unremarkable}
Tolmie et al.~\cite{unremarkable-computing} introduced the notion of unremarkable computing when discussing how ubiquitous computing should arrive and create its place in people's homes. They argued that technology can augment people's actions in ways that have a wealth of significance but seem unremarkable, because its interactions are ``so highly situated, so fitting, so natural''. They argued that home technology should not only be more intelligent, it should also be more subservient to people's daily routines. In doing so, the technology becomes part of the routines, part of the very glue of their everyday life.

We draw connections between this ambition and our aforementioned design goals.
We also draw connections between this notion of routine and VAD decision making. While these are daunting life-and-death decisions, the implant decisions are part of a work routine for clinicians. To fit a DST into their practice, we need to make it subservient to the day-to-day decision-making workflow they engage in. 

We wanted to operationalize this idea of unremarkable technology in the context of critical, clinical decision making. This is a difficult goal because it requires a right level of ``unremarkableness" such that the DST does not constrain clinicians' decision making flow \textit{except when it needs to}. 

\subsection{Design Process}

To situate a DST into the current VAD decision-making routine, we first needed to identify a time and place where clinicians should naturally and impactfully encounter the it. 
We chose the multidisciplinary patient evaluation meetings, for a number of reasons. First, the meeting is a rare social touch point where most clinicians involved in the decision are present, and they are actively forming a collective decision about patient treatment. Second, it is one of the few decision points where a computer is present and being used. Third, decision meetings are common across hospital sites. VAD centers in the US are legally required to take a multidisciplinary approach to patient care, therefore regularly scheduled meetings are common. Globally, these meetings are also recommended \cite{multidiscplinary-meeting-VAD}. Fourth, multidisciplinary meetings have become an increasingly common best practice in organ transplantation \cite{tumor-board-effective}. Designing DST for decision meetings therefore could potentially generalize beyond VADs to include a number of other clinical decisions.

Next, we considered how to fit the DST comfortably within the meetings. Drawing lessons from prior work \cite{o2007toward,yangVAD2016}, we wanted to embed the DST into Electronic Medical Records (EMR) to minimize the effort needed from clinicians to type in patient information. We also wanted to augment clinicians' paperwork to provide them additional motivation for adoption. We therefore integrated the DST output into a meeting slide generator, a system that automatically extracts patient information from EMR and populates slides for the decision meeting, which could be projected or printed.

We sketched what the DST predictions output might look like. We iterated on the design based on feedback of two collaborating clinicians (an attending Cardiologist and a nurse practitioner). The final design was a small line chart that showed a patient's predicted chance of survival (Figure \ref{fig:screenshot}). It also showed the most likely causes of death, such as right ventricular failure or renal failure.

We placed this chart in the top-right corner of the slide summarizing an individual patient's current state. The subtlety was a deliberate choice toward achieving the right level of unremarkableness. In the most common case, when the DST agreed with the clinicians' assessment, the visual display of the agreement could help clinicians gain trust in the system without slowing them down. In the rare case that the DST prediction conflicted with the clinicians' assessment, the DST could slow the decision down. Everyone attending the meeting would see the disagreement. We speculated this would apply social pressure on the senior physicians to rationalize and articulate their decision making. We speculated it could also encourage the medical students, residents and other mid-level clinicians to participate in the discussion when they disagreed with the senior clinician's decision. It could allow them to disagree by pointing to the conflict with the DST and not claiming that they personally knew more than the senior physician.

We worked out the detailed contents of the slide with the two collaborating clinicians. We also referenced the meeting printouts and workup checklists currently in use. 

We wanted to finalize the design by populating with real patient data. However, a variety of policies and legal regulations would not allow this. As a work-around, we asked our clinical collaborators to help us populate the slides with synthetic patient cases. Interestingly, they found it very challenging to generate a prototypical patient case including dozens of vital signs and test results. They instead selected elements across several of their former cases, removing identifiable demographic information and molding parts of the medical condition to disguise the identity. 

In our final design (Figure \ref{fig:screenshot}), the DST outputs are in the top right corner of the slide, next to a summarized patient history visualization. Patient test results are categorized and put in the center. The patient demographics and links to social and financial evaluations are on the left.

\section{Design Assessment}

We had several questions we wanted to answer with our assessment, including: (1) Would clinicians naturally encounter the DST within their current workflow? (2) Would clinicians accept computational decision support in the public context of the meeting? (3) Does placing the prediction in the corner present the right amount of unremarkability? Specifically, does the DST get ignored when its predictions align with the clinicians' judgment, and would it slow decisions down when its output conflicts with clinicians? 

\subsection{Assessment in VAD Implant Centers}
We gained access to three hospitals that regularly perform VAD implantation, all within the US. 
Two were sites from our formative field study and one was new. 
The facilities varied geographically and in scale. The smallest we studied performs about 40~VAD implants a year; the largest performs over~100.

We wanted to assess our design within the context of an actual implant decision meeting in order to observe whether it impacted discussion. Unfortunately, this proved to be impractical. None of the sites would allow us to present slides showing information for the patients they were currently implanting. All felt this could impact the life and death decision. The clinicians doing the VAD implants were quite busy. They would only agree to interact with a single design. They did not have the time for us to make revisions and then revisit. Finally, one of the sites had a specific policy preventing us from observing the decision meeting. They would only participate in one-on-one interviews.

In reaction to these restrictions, we re-designed the assessment process with the goal of making the most use of our participant pool within one round of assessment. We carried out all following procedures in hospital C. In hospital B, we carried out all except (3) presenting at a decision meeting. In hospital A, we carried out all procedures except (4) interviewing all physicians and surgeons.

(1) At each site, we first interviewed the mid-levels to understand their practice around the decision meeting, and to probe the DST design's fit in their respective hospitals. When necessary, we adjusted the designs to fit specific hospital's routine practice;

(2) Our research collaborator at each site recommended one attending physician to be our confederate. We conducted interviews with them, discussing the DST design, and confirming there was no glaring mismatch between the design and the practice at their respective sites; 

(3) The confederate physician presented the patient case with the DST on display in the decision meeting. We observed clinicians' responses and discussions;

(4) Finally, we interviewed the rest of the VAD team to further individually discuss the DST design.

In total, we interviewed nine attending cardiologists or surgeons and eight mid-level clinicians. Each interview lasted for at least one hour. The DST design was presented in two hospitals' multidisciplinary decision meetings. Field notes were recorded using pen and paper. Interviews were audio-recorded and transcribed. We analyzed our data using affinity diagrams \cite{moggridge2007designing} and by performing thematic analysis.

\subsection{Assessing Generalizability of the DST Design}
We chose to situate the DST within slides used for decision meetings partially because these meetings are best practices in other critical medical domains as well. To gain some insights as to if this design might generalize, we chose to probe a small set of clinicians from other medical domains who participate in these meetings.

To recruit these participants, we asked participants from the VAD study to help us identify other clinical domains and decisions that have interdisciplinary decision meetings. We then interviewed 6 physicians whose practices include decisions meetings for pediatric surgery, pediatric critical care, adult cardio-thoracic surgery, internal medicine emergency care, orthopedic surgery, and obstetrics/gynecology. We audio-recorded, transcribed and analyzed these interviews using the same methods as we used for our VAD participants.

\section{Design Assessment Findings}
We first offer an overview of observations from the individual sites, describing the different cultures, facilities, and practices. We then report findings across the three sites related to the aforementioned assessment goals: the likelihood of encountering DST during decision-making, the acceptance of DST, the right level of remarkableness, and finally, generalizability to other kinds of medical decisions.

\subsection{Overview}
Hospital A was the least technologically advanced. They recently transitioned from paper-based to electronic clinical records. Phone signals generally did not penetrate the building, built in the late 1800s. Many common web services, such as Google search, were blocked on their internal network. 

The weekly meeting took place in a long, grandiose, turn-of-the-century board room. This contained one long, 40-seat table above which hung four large chandeliers. At one end of the table (the ``head'' of the table) there was a portable, low resolution projector. They sat according to an unspoken seat chart based on clinical role hierarchy. Physicians sat near the head end, and a small group of these physicians would present the individual cases. Nurses sat near the middle. Social workers and others sat farthest from the projector. Only participants sitting near the head of the table participated in the discussion. The content on the projector screen was also too small to read since most sat far from the screen. They all interacted with bulky printouts of EMR data for each patient discussed. 

Hospital B had in-house statisticians dedicated to outcome analysis and patient risk modeling. The physicians and this site were also actively involved in VAD risk modeling research. Interestingly, when it came to using a risk model to inform their own implant decisions, they described themselves as ``\textit{very minimalist despite all these interests in ML.}'' Cardiologists and surgeons led implant decision making both within and outside of the implant meetings. Meeting participants did not vote on how to proceed. Hospital B did not provide us authorization to observe its decision meeting. 

Hospital C was more technology-friendly. The meeting room had large projector, which most participants could read. In addition, participants had access to a printout of the presented materials. One program manager and two mid-level clinicians arrived more than 40 minutes before the meeting to set up the computer, projector, and remote conference connections. As the presenting physicians spoke, a seasoned nurse practitioner operated the computer, pulling out and zooming into relevant patient information from EMR. This nurse practitioner had been performing this role for more than 5 years. Physicians and mid-levels used laptops to search for relevant information in the EMR or online and to add items to their digital to-do lists. Many more people engaged in discussing the patients. Following the discussion of each patient, all clinicians present voted on the next step.

Hospital C had previously experimented with bringing computational predictions into their meetings. Cardiologists chose a model that had been nationally validated through five randomized clinical trials. They had a nurse practitioner input all of the data for each patient discussed and show the DST prediction in the decision meeting. One year later, they stopped this practice because two recent journal articles reported that the models used were ``\textit{horribly mis-calibrated}''. ``\textit{That was a lot of work to type in all that sh-t and generate that number, and that's not that helpful.}'' Their EMR held four other implant outcome prediction models, which predicted things such as the chance of depression. However, the clinicians never used these models, stating that each required manually entry of all of a patient's data. 

\subsection{Likelihood of Encountering DST in Workflow}
Our observations suggested that most clinicians involved in the VAD implant decision would likely encounter the DST output if it was included as part of an individual patient's information presented at the decision meeting. All three facilities hosted a weekly implant decision meeting. Clinicians of all ranks and roles attended, ranging from seasoned surgeons to residents, to nurse practitioners to social workers to palliative care coordinators. Although the weight that the meetings carried for influencing an implant decision appeared to vary across the three sites, the occurrence of the meetings was one of the few events that happened everywhere. 

These meetings offered one of the extremely few situations where senior clinicians actively discussed decisions in proximity of a computer. Meetings in all three hospitals had a shared computer projecting patient information. Two hospitals projected dedicated meeting materials. The other projected patient profiles from the EMR. Clinicians described the other key decision points as ``just talk on the fly'' with no EMR access or paper records in hand. The other decision points most often only included attending physicians and surgeons. ``Everything is happening live.'' Mid-level clinicians, who spend more time with each individual patient, did not participate in the decisions made outside of the meeting.


\subsection{Acceptance of DST in Decision Meetings}

None of our interview participants expressed any resistance to the including DST output within the context of the decision meeting. One site (Hospital C) had already made the effort to manually include DST data into their meeting but had abandoned this practice due to their loss of confidence in its quality. Seasoned physicians and surgeons voiced their appreciation for what a prognostic DST might bring, stating that it would ``\textit{give its perspective}'' and offer a chance for an ``\textit{occasional recalibration.}'' Clinicians also shared that making an objective decision could sometimes be hard. The decision to not implant was usually a death sentence for a patient. ``\textit{When I really like this patient, really want to help him or her, it sometimes helps to get a more factual view.}''

Seasoned physicians shared that their dream DST should play a role similar to mid-level clinicians. They should provide additional context for the seasoned physicians' decision. The DST could provide additional context and a different perspective to the senior physicians. They recognized the value a DST might bring from its statistical consideration across many cases. ``\textit{The value is you are looking at thousands of cases, I'm looking at 100 and overweighting the last three I saw.}'' They also shared that input from mid-levels was not always ``\textit{taken really into account}''.

Mid-levels agreed they only inform and support the discussions. They did not make decisions.

\begin{myquote}
\textit{My role in selecting patients for VAD... hmm. I don't select patients. But I do talk about it... We are there to help discuss patients.} (Nurse practitioner, B2)
\end{myquote}


Mid-level clinicians enthusiastically welcomed the idea of a decision meeting slide generator. They envisioned a number of possible benefits. They shared that the slide generator would automate work that is not currently billable. At hospital A and B, meeting slides were prepared by staff who had little to no medical training. Physicians could get frustrated with the result, characterizing the unfiltered materials as being prepared by ``amateurs.'' These staff members could not personalize patient presentations because they could not risk skipping information that might prove to be critical. Mid-levels felt they could benefit from the automation and seasoned physicians felt they would benefit by the removal of the copious, irrelevant data being pulled out of the EMR. 

Mid-level clinicians viewed the slides as a potentially important vehicle for communicating their opinions to physicians. In all three hospitals, senior physicians set the agenda for decision meetings. They decided which patients to present, and during the meeting, they called out the information that they felt was important enough to discuss. This hierarchical culture was well captured by the design of a custom patient review tool at hospital C. Two VAD coordinators customized a patient review dashboard within EMR in order to help themselves better track medical tests and share results within the team. Although cardiologists and surgeons rarely used the tool, they controlled which pieces of information could be placed on the dashboard and which elements would not be included when the patient case was classified as urgent. 

Mid-levels often doubted that their voice was heard or that their expertise was considered. They were hesitant to directly disagree with a physician. They described the situation as more complicated than just the power dynamics. They shared that the cardiologists were incentivized to implant more patients and to implant sicker patients. They found themselves often advocating for patient mortality (let the patient die). Mid-levels felt their opinions focused on post-implant quality of life. Unlike the physicians, mid-levels worked intimately ``\textit{with all the problems that can come from a patient that maybe shouldn't have been implanted.}'' They noted there was no right or wrong answer between length of life and quality of life. They shared it was often hard to argue with great confidence that letting patients die was better than offering them a small chance to live. In such situations, mid-levels frequently cited ``\textit{you never know what will happen}'' as a reason to not to pursue further discussion with attending physicians. Some shared that over time, they had slowly removed themselves from the decision making. 

\begin{myquote}
    \textit{There is risk stratification for each patient, but I don't know... It's like, we talk about it, but I don't know if it's really taken really into account. } (Nurse practitioner, B2)
\end{myquote}

Mid-levels consider the ability to organize the contents of meeting slides as one way to increase their influence. Meeting slides provide additional, visual presence they could use in support of the facts they felt were important. This would make it less like they were only sharing an opinion with the physicians. The meeting slides could be facts in a space where only the seasoned physicians' opinions carried any weight. They felt the formality the meeting slides carried was unparalleled to any other artifact they had access to. A prognostic DST that indicates post-surgery quality of life could potentially amplify their voices.

\begin{myquote}
    \textit{There is not a way to present (my reasoning) formally. It's just me saying: `This, this and this'. [...] I think it's good to have something visual for anybody to see. It's like, OK. LOOK. Let's slow down a bit here.} (Nurse practitioner)
\end{myquote}







\subsection{Intricacies of Making DST Unremarkable}

Both seasoned physicians and mid-levels expressed appreciation for DSTs that could slow them down ``only when necessary". They liked this aspect of our design. 
However, we could not easily conclude whether our specific design had achieved this goal. Instead, clinicians' discussions and questions, which we will soon describe, depicted many unexpected intricacies in this notion of the ``right" level of unremarkableness.

\subsubsection{Challenges of Engaging Synthetic Patient Cases via Data}
Clinicians shared that they could not draw on their experience of making critical clinical decisions seeing only patient data on paper. This presented the biggest barrier to assessing how clinicians might respond to a conflicting DST prediction. 

Patient history data alone did not give clinicians enough confidence to make an implant decision. Physicians described the meeting data as merely a surrogate for the actual patient. The data did not allow them to see patients ``\textit{as a whole.}'' They stressed that to understand a patient clinically, they needed to ``\textit{look at the patient, talk to the patient, take care of the patient.}'' Social workers shared that they had not met with this patient nor talked to their family. In our field evaluation, presentations of the synthetic patient cases were always followed by a long, awkward silence.

\begin{myquote}
\textit{A very sick but highly motivated patient can do better than their illness would otherwise be left them, compared to a less sick, less motivated patient. These things are hard to capture. The eyeball tests.} (Surgeon, B6)
\end{myquote}

Clinicians also had wildly different readings into the same DST prognostics. We presented the same two synthetic cases with the same implant survival predictions to all participants. Interestingly, they generated wildly different reactions and interpretations of the cases. Some viewed the survival estimate as implying that an implant would not work. ``\textit{Gee...~VAD is futile here.}'' Others viewed the DST output as implying the patient should be immediately implanted, before things got worse. `\textit{`We still have a chance.}'' Few clinicians believed that all VAD implant candidates would have a similar prognosis as the synthetic case we presented: ``\textit{This chart is meaningless. Every VAD candidate's projection would look like this.}'' 

That the data was based upon synthetic patient cases made any real discussion about the patient even more difficult. Instead, clinicians started to focus on the DST prognostics. They probed on where the model comes from. It took a long time for us to explain the data source and the ML mechanism to clinicians with no ML experience and without a deep understanding of statistics. It took even longer to explain it to clinicians with statistical depth and ML experience. They fixated on the fact that the ML systems' performance was not the focus of our assessment. The synthetic patient data often turned this into an assessment of the DST's quality in the minds of many meeting participants.

\subsubsection{Is the Model Validated by Clinical Trials?}

Clinicians commonly expressed a need to know more about the model's source and credibility. When they learned that the model presented has not been rigorously validated through clinical trials and published in prestigious clinical journals, they suggested we were wasting their time. Physicians and surgeons considered discussing an unvalidated model unethical; as misleading as ``\textit{looking at a crystal ball}''. Others tended to judge DST quality based on the journal it was published in. 

Physicians also desired a model that had been validated with data from their own hospital. ``\textit{It's better to be home-grown.}'' Models should be published in a good journal and then validated in a national scale study across several implant centers. Some suggested including links to the peer reviewed clinical trial within the DST output on the slide. It ``\textit{lends a lot of weight to a clinical model}''.

\subsubsection{Are the Predictions Based on Clinicians' Best Efforts?}

Physicians highlighted that the predictive models, regardless of how well they measure medical uncertainties, would never replace human, clinical decision-making. They viewed their own decision making as focused on managing and reducing uncertainties. ``\textit{If we think that we will be able to tell everybody what to do based on a model, we ignore the fact that we also have tools and mechanisms for dealing with the uncertainty that is inherent when putting VADs in patients.}'' (Cardiologist)

Many clinicians' questions, as well as their discussion around the DSTs, revealed a tension between what they saw as the DST's static view of patient conditions and the clinicians' desire and ability to also focus on future actions and interventions. They wanted to know which modifiable factors most influenced the DST predictions. They wanted to be able to offer treatments that they could improve these factors, thus increasing the likelihood of a positive surgical outcome at some time in the future.

\begin{myquote}
\textit{These predictions are (what will happen) despite our best efforts, right?} (VAD manager, C8)
\end{myquote}

\begin{myquote}
\textit{Having an understanding of what's driving the risk [features that most influence the prediction] is very important for us to understand what is modifiable at that patient. [...] Is it age or something we cannot change? Otherwise there is a lot of potential here.} (Hospital C decision meeting)
\end{myquote}


Clinicians did not seem to actively make the subtle but critical distinction between features that were important to predicting an outcome and features that are causal to that outcome. For example, an observation that people are carrying umbrellas can be used to predict that it will rain. However, taking people's umbrellas away will not prevent rain. ML systems make predictions based on covariance of features. They do not assess the causality of those features. When prompted, clinicians claimed that this distinction is ``\textit{absolutely important}''. However, in our conversations, we did not observe them distinguishing ML predictions from general statistics. They seemed to strongly believed DSTs should be able to distinguish causality from prediction and that they should present only causal features. ``\textit{This is the whole point of statistical processes. A DST model should address that, right?}"

There was a sense that if the DST predictions were not based on causal factors, then the predictions should not be presented at all. Clinicians described differentiating correlation (predication) versus causality as a central part of their decision making. For example, many patients being evaluated for left-ventricular VAD also have right-ventricular heart failure. An important decision cardiologists must make is whether the heart failure on the right was caused by the left heart failure or if it is independent. Will fixing only the left side also fix the right? Currently, clinicians speculate by probing patients with medication. They try different left heart medications and observe how the right side responds. Clinicians wanted help: ``\textit{If you can help us understand [...] which factors seem to be most dominant, or most closely associated with certain outcomes, then that helps.}'' They wanted to know the causal links and features for individual cases. 

\subsubsection{Are Data-Driven Prognostics Facts OR Predictions?}

Clinicians frequently asked us to clarify whether DST prognostics are predictions that carry agency and subjectivity, or if predictions are facts rooted in historic data. We sensed they wanted to limit discussions to facts, including how heart failure has played out for the patient they were treating and the statistics from previous, similar cases. We observed resistance from some clinicians toward the idea of showing predictions. Our collaborating physicians, who created the synthetic cases and helped us select contents for the slides suggested that the DST output should be ``\textit{one statistical representation of 100 patients who are similar to him}'' rather than a prediction for this individual patient.

\begin{myquote}
\textit{I think if you continue to call it ``VAD projections'' 65\%, people are going to poke holes at it. They are gonna try to prove you wrong. This [DST projection] is just what the historical outcomes were. But this guy is different, this guy has his own things that make him special.} (Collaborating cardiologist, hospital A)
\end{myquote}

\subsubsection{Are the Predictions Individual Medicine OR Population Medicine?}

Most clinicians share that they thought of DST output as an ``\textit{average}''. They seemed to find the notion of personalized predictions difficult to grasp. Some voiced strong concerns that using DST was the same as applying ``\textit{populational statistics}'' to individual patient decision making. They felt this was unethical. Others proposed that ``\textit{instead of having one model that we apply to the entire population, we would have a group of models. Those models predict for that group of patients.}'' (Surgeon, B4)


\subsubsection{What Does ``Now'' Mean in DST Predictions?}
The DST visualized the patient outcome predictions, including life expectancy, estimated time until right heart failure, and likely cause of death. For example, Figure \ref{fig:screenshot} shows that the patient's post-implant life expectancy is 21 days if a VAD was implanted now, under the condition shown on the slides.

Clinicians were confused by this notion of ``now'' because it was extremely unlikely that they would implant a patient on the same day as the decision meeting. Is ``\textit{that 21 days from today? If we are gonna lose the patient in 21 days [21 days following after implant], can we just wait?}''

\subsubsection{DSTs Do Not Account For the X Factors}

Clinicians said that the DST would only ever be one factor in their decision because of ``\textit{X factors}''; the many factors beyond a patient's condition that impacts the implant decision. One X factors they spoke of was O/E ratio (observed-to-expected mortality ratio). The O/E ratio is a rating that measures the surgeon and care teams' performance. Surgeons cared about keeping a high rating. They described the implant decision for high-risk patients as ``\textit{taking on new O/E ratio debts.}'' This seemed to strongly influence whether they take on another high-risk patient. It seemed to depend strongly on how many patients had recently had poor outcomes. 

\begin{myquote}
\textit{It's not that we don't help that [VAD candidate] patient, but if we take this shot and do poorly, then we cannot take on the next 10 patients like him. Because now we got too much of a cluster of high-risk patients who've done poorly, then we have to do some lower risk ones before we can go back up [in O/E ratings]. Insurance companies and Medicare and all that... they will mark you. They may not pay. It all plays into the complex factor for deciding who, especially sicker patients, we would take a shot.} (Surgeon, B6)
\end{myquote}

Some surgeons described that, for some cardiac surgeries that have officially defined models used to rate surgeons and care teams, their decision meetings had became centered around risk models. This is not yet the case for VAD implants.

\subsection{Generalizability Beyond VAD}

Our interviews with clinicians outside of VAD centers showed that multidisciplinary decision meetings take place across many clinical domains for some of their most aggressive interventions. They are also referred to as internal medicine panel meetings, tumor boards, or floor meetings (referring to meetings between critical and general care physicians). These meetings happen widely because for patients are very sick and are being considered for their last-option surgical intervention, their illness usually have involved multiple organs. Treating them requires physicians from multiple clinical domains. Multidisciplinary meetings therefore occurred naturally.  

\begin{myquote}
\textit{Esophageal cancer, COPD, diabetes, cystic fibrosis, LITERALLY everything in psychiatry, gastric bypass, end stage renal disease, hernia repair, syndromes like Down and Turner, any disease that requires management with meds with nasty side effects, and even emergency room situations to expedite processes. Any of the above diseases the approach has to be multidisciplinary almost by definition because they affect multiple systems and usually but not always the last option is a surgical intervention.} (Pediatric surgeon)
\end{myquote}

\section{Discussion: Designing and Evaluating DST as A Situated Experience}

Clinical DSTs, despite compelling evidence of their effectiveness in labs, have mostly failed when moving out of labs and into healthcare practice \cite{sittig2008grand,taneva2014meaning}. A lack of contextual integration in the design of these systems played a critical role in these repeated failures. Prior work suggests that current interaction conventions, that clinicians will recognize their own need for a DSTs help and then walk up and use a system separate from the EMR, is not likely to work \cite{yangVAD2016}. 

There is a real need to design DSTs not only as a functional utility but as an integrated experience. Their effectiveness should be measured not only by prediction accuracy, but by effectiveness when situated within its social and physical context such as workplace culture and social structures. This presents exiting new opportunities and challenges to HCI and UX research.

Our design makes three dependent proposals about making a DST a situated VAD decision making experience. First, we propose that the decision meeting presents a good \textit{time and place}. Second, assuming the meeting is correct, we propose that situating the DST output into the meeting slides would offer an effective \textit{form}. Third, assuming that having the DST as part of the slides is a good form, we propose that the DST plays a \textit{fairly unremarkable} role in clinician decision making by appearing in one corner. We claim it needs to be easily passed over when it agrees with current decision making and that it must only be present enough to slow decision making down when its predictions are in conflict with a seasoned physician's suggested course of action. All three proposals aimed to naturally augment the current activities of decision making, rather than pulling clinicians away from doing their routine.

Below, we discuss the design implications of these proposals. We then share challenges encountered and lessons learned in evaluating the DST as a situated experience.

\subsection{Designing DST to Augment Clinical Routine}

\subsubsection{Time and Place}
Findings of this work suggested that DSTs may more effectively fit into clinical practice if their interactions are tailored for a specific time and place within the current decision-making workflow. Taking lessons from prior HCI work, we should not only make AI more intelligent, but make them highly situated in people's routines. In doing so, AI can become part of the decision-making routines, part of the very glue of clinicians' everyday work.

Our assessment findings largely suggest that decision meetings are a routine activity that is promising for DST integration, for several reasons: 

\begin{enumerate}[noitemsep,leftmargin=*]
\item The meeting is part of an existing clinical decision-making \textit{routine}. Clinicians therefore would naturally encounter the DST at the meeting;
\item The meeting is a \textit{socially aggregated} decision point. The DST could therefore leverage mid-level clinicians to advocate for its information and value to the decision makers;
\item The meeting offers a moment of \textit{deliberation} in their otherwise fast-moving decision-making workflow. The meetings offer clinicians time to collectively digest the implications of the prognostics; 
\item Finally, the meeting is a best practice promoted globally in VAD patient care, and across several clinical domains. Therefore this DST design could potentially make its place across diverse practices in different hospital sites and domains.
\end{enumerate}

Decision meetings represent only one way of integrating DSTs into clinical practice. Similar opportunities may lie in other time and place in existing clinical decision-making routine that is socially-aggregated, deliberative and shared across hospitals. Future research shall advance this work by systemically searching for such opportunities. 

\subsubsection{Interaction Form}
Besides situating the DST in decision-making routine, we also motivated mid-level clinicians' use by preparing patient information for the decision meetings for them. Our field study suggested this was a useful tactic. DSTs supporting various clinical decisions can potentially automate tedious information retrieval tasks for clinicians to offer additional motivations for adoption.

\subsection{Designing a Right Level of Unremarkableness}

The walk-up-and-use convention of current DSTs assume clinicians will know when they need help. Our design challenges this convention by proposing the notion of \textit{Unremarkable AI}. Our unremarkable DST is designed to be situated naturally in an existing decision-making routine and only noticed when it might add value to the decision. DST's interaction should have a right level of unremarkableness, yet the information it provides should significantly impact care. 

Our field assessment illustrated \textit{some} positive indicators that making DSTs unremarkable helped reduce the resistance clinicians commonly show towards clinical DSTs \cite{yangVAD2016,gravel2006barriers,taneva2014meaning}. For example, we did not observe clinicians feeling threatened or feeling they might be replaced by the technology. Clinicians appreciated that DSTs could inform their discussions, ``\textit{though the discussion is unlikely to center around the DST.}"


While our DST was visually unremarkable, its very existence seems be, to an extent, transforming clinical decision making. It introduced predictions into a culture rooted in facts and statistical significance. Moreover, when predictive risk models were used officially to measure patient risk and clinician skills, clinicians' decision making became centered around these models. DSTs substantialized their performance pressure in decision making.

These observations forced us to take a step and ask: What is the preferred role for DST to play in clinical practice? Where does a right level of unremarkableness lie? More research is needed to find the right balance between DST \textit{augmenting} decision-making in natural and intuitive ways and \textit{transforming} the nature of clinical decision-making. Understanding these tradeoffs should be a critical research question in DST design and research. 


\subsection{Experience Prototyping DST In-Situ}

Restricted access to the clinical environment is known to impose fundamental challenges to iterative UX design and evaluation. Our experience of conducting the field assessment echoed this. Upon reflection, we identified several tactics effective at reducing the risks of our one-shot design evaluation:

\begin{enumerate}[noitemsep,leftmargin=*]
    \item Designing a generalizable DST: The work flows and social contexts in clinical practices are complex and highly divergent across hospitals. Therefore, generalizability is a necessity for many DST designs. This work took a step further than hospital-site generalizablility, designing a DST that can work for a class of structurally similar decisions (data-intensive, last-option surgical interventions). A DST's design and evaluation can become more productive than those dedicated to one specific clinical decision as well as specific DST models;
    \item Designing the evaluation methods for describing and unpacking the complex, subtle, and multi-faceted nature of experience, rather than explicitly measuring it;
    \item Using prototypes rather than functioning DST models. This allowed us to probe various possible DST outputs and to easily adjust our prototype to incorporate participant feedback.
    
\end{enumerate}

\subsection{The Impossibility of \\Experience Prototyping Critical DSTs}
Nonetheless, we encountered additional, seemingly-inevitable challenges of assessing DST's situated user experience. For example, whether a DST design has indeed achieved a right level of remarkableness was impossible to assess without real patient data and fully functioning ML systems. 
 
Clinicians need more than just synthetic patient cases to connect with their own decision making. We speculate that clinicians need to see one of their own patients' data to really assess the DST information design and to see what an actual prediction would look like. This means early DST prototypes will need actual patient data to assess their interactions in context and their impact on care. This is currently impossible in critical clinical cases due to ethics, policies and hospital regulations. 

Clinicians were unable to engage in a group discussion without a fully functioning ML system. Clinicians described using an unvalidated DST as unethical and misleading. They suggested that a DST should be validated via randomized clinical trials on both retrospective patients and \textit{prospective} patients, both at a national level and on their own hospital's patient population. This gives rise to a chicken-and-egg problem in our design assessment: Clinicians could not effectively assess the DST design without a working DST that has been validated on prospective patients; and validating a DST on prospective patient data requires a DST design that has been proven effective. 

We suspect these challenges are likely to occur not only in evaluating DSTs for artificial heart implant, but in assessing DSTs for many other critical, high-consequence decisions as well. As data-driven DSTs increasingly move out of research labs and into critical decision making in the real world, we encourage DST designers and researchers to join in making these challenges explicit and investigating new design assessment methods and tools to address them.

\section{Acknowledgement}

This work was supported by grants from NIH, National Heart, Lung, and Blood Institute (NHLBI) \# 1R01HL122639-01A1. The first author was also supported by the Center for Machine Learning and Health (CMLH) Fellowships in Digital Health. We thank the participants in this work for their dedication, time and valuable inputs.

\bibliographystyle{ACM-Reference-Format}

\balance
\end{document}